\documentstyle[preprint,eqsecnum,aps,tighten]{revtex}
\input epsf.sty

\newcommand{\be}{\begin{equation}}
\newcommand{\ee}{\end{equation}}
\newcommand{\bse}{\begin{subequations}}
\newcommand{\ese}{\end{subequations}}
\newcommand{\bea}{\begin{eqnarray*}}
\newcommand{\eea}{\end{eqnarray*}}
\newcommand{\ba}{\begin{array}}
\newcommand{\ea}{\end{array}}
\newcommand{\p}{\partial}

\newcommand{\dfrac}{\displaystyle\frac}

\newcommand{\bom}{{\mbox{\boldmath$\omega$}}}
\newcommand{\bOm}{{\mbox{\boldmath$\Omega$}}}

\newcommand{\bPsi}{{\mbox{\boldmath$\Psi$}}}

\newcommand{\bmu}{{\mbox{\boldmath$\mu$}}}
\newcommand{\bm}{{\mbox{\boldmath$m$}}}
\newcommand{\br}{{\mbox{\boldmath$r$}}}

\newcommand{\bk}{{\mbox{\boldmath$k$}}}

\newcommand{\bp}{{\mbox{\boldmath$p$}}}
\newcommand{\bu}{{\mbox{\boldmath$u$}}}
\newcommand{\bU}{{\mbox{\boldmath$U$}}}

\newcommand{\bQ}{{\mbox{\boldmath$Q$}}}
\newcommand{\bH}{{\mbox{\boldmath$H$}}}
\newcommand{\bj}{{\mbox{\boldmath$j$}}}
\newcommand{\bJ}{{\mbox{\boldmath$J$}}}

\newcommand{\cas}{\textstyle\frac}
\newcommand{\bnabla}{{\bf\nabla}}
\newcommand{\upi}{\pi}

\newcommand{\ve}{\varepsilon}

\newcommand{\dd}{{\rm d}}
\newcommand{\ir}{{\rm i}}

\def\etal{\mbox{\it et al.\ }}
\begin{document}
\title{The evolution of three-dimensional localized vortices
in shear flows. Linear stage}
\author{I. G. Shukhman}
\address
{
Institute of Solar-Terrestrial Physics, Russian Academy
of Sciences, Siberian Branch,\\
Irkutsk 664033, P.O.Box 4026, Russia
}
\author{V. Levinski}
\address
{Faculty of Aerospace Engineering, Technion - Israel Institute of
Technology,
\\Haifa 32000, Israel\\
(Present address: KLA-Tencor Corporation,
Migdal Ha Emeq 23100, Israel)}

\maketitle
\begin{abstract}
{\footnotesize The evolution of a
small-amplitude localized vortex disturbance
in an unbounded shear flow with the linear velocity profile is
investigated. Based on the exact solution of the initial problem,
a revision is made
of the theoretical approach (suggested by Levinski 1991 and
subsequently further developed in a series of other publications)
in which the vortex evolution is described in terms of Fluid Impulse
of the vortex ``core". Although the theoretical predictions obtained
on the basis of that approach were excellently confirmed in subsequent
experimental studies, its inconsistency is demonstrated in this study.

    According to this solution, the localized vortex increases
slowly (as power-law with the time) and attains an almost  ``horizontal"
orientation, unlike the previous theory (Levinski 1991) that predicts
the more rapid growth and vortex orientation at the angle of
$45^{\circ}$ to the flow direction.
On the other hand, just the rapid increase and
the angle of $45^{\circ}$ to the outer flow direction are characteristic
for hairpin vortices observed in turbulent boundary layers or
artificially synthesized vortices in laminar boundary layers.

    Thus the issue of adequate theoretical interpretation of the evolution
of localized vortices is again on the agenda.
The remaining part of the paper presents the
first steps in the solving this problem.
In particular,
the dynamics of the total enstrophy of vortex
as the measure
of vortex intensity
is followed. The dependence of
vortex amplification on its initial orientation
is investigated.
On this base the validity of the old
idea of Theodorsen (1952) on the predominant formation
of the $45^{\circ}$ vortices is discussed.
Also the tensor of enstrophy
distribition (TED) is defined and it is shown that
it may serve an effective tool
in describing of the vortex geometry.

The linear stage of Gaussian vortex evolution presented here
provides a very suitable base for testing of further numerical simulation
of the nonlinear stage.
}
\end{abstract}

%
\section{Introduction}
Two main types of coherent vortex structures, which were first
identified using flow visualization technique in experiments
reported by Kline \etal (1967), form the basis of the present
views of the structure of a turbulent boundary layer.
Thus the presence in the wall-bounded  flow of streaks,
along which the streamwise velocity is lower
than the average velocity at the same distance from the wall,
results from the rise of low-velocity
fluid from near-wall layers induced by long-lived vortex
structures. These latter  represent a pair of counter-rotating
vortices extended along the flow direction
(Bakewell \& Lumley 1967; Smith \& Schvartz 1983).

     Another phenomenon, widely observed in turbulent
boundary layers and referred by Kline \etal (1967) as
``bursting", results from the rapid evolution of the localized
vortex having the shape of a hairpin. These vortices were found
to be inclined at $45^{\circ}$ to the flow direction (Head \&
Bandyopadhyay 1981), and their typical lifetime are about 5\% of
the streak lifetime.

The evolution mechanisms of these well-organized vortex structures
and their interaction have been subjects for
study by an ever increasing number of researchers
(see reviews by Robinson 1991 and Smith \& Walker 1983).

     Despite the fact that both types of coherent vortices
have an identical structure of the type of vortex dipole,
their properties, and also formation and
development mechanisms differ greatly.
The slow evolution of near-wall vortices is reasonably well
explained by the mechanism of algebraic growth suggested by
Benney \& Gustavsson (1981)
and subsequently further developed in terms
of the concept of optimal disturbances by
Butler \& Farrell (1992), Reddy \& Henningson (1993) and
Reshotko \& Tumin (2001). On the other hand,
the derivation of
an adequate theoretical model describing the evolution
of hairpin vortices is complicated by
the high degree of vorticity
localization  in the core of the hairpin
vortex and, hence, by  the  strong
nonlinear character of their development
from the outset. This is confirmed
by a number of experiments where
hairpin vortices were observed only at the transition
of the threshold value of a certain parameter corresponding
to the mechanism of their generation used in the experiment.
Thus in experiments of Asai \& Nishioka (1995)
where hairpin vortices were generated by using
acoustic disturbances, these vortex structures were
observed only when the amplitude
of the applied  disturbance stood out above a certain critical
value.
In experiments of Malkiel, Levinski \& Cohen (1999)
the initial disturbance was created by employing suction
of fluid through the holes in the wall.
Here, as in the previous case,
hairpin vortices were observed only
at fluid suction rates exceeding a certain critical value.

     Based on the aforementioned factors, it is of interest
to analyze the theoretical model describing the evolution
of a nonlinear localized vortex disturbance in the external plane
shear flow first suggested by Levinski (1991)
and subsequently generalized to rotating flows (Levinski \&  Cohen 1995),
flows of weakly conducting fluid in a magnetic field
(Levinski, Rapoport \& Cohen 1997) and to stratified flows
(Levinski 2000). In this model the vorticity distribution
is characterized by its fluid impulse integral  defined as
$$
{\bp}={\cas{1}{2}}\int{\br}
\times \bom(t,{\br})\,\dd V,
\eqno{(1.1)}
$$
where ${\br}$ is the position vector,
$\bom(t,{\br})$ is the instantaneous field of
vorticity disturbance, and the integration is done over the
entire volume of fluid. Accordingly, the fluid impulse
dynamics is described by the equation
$$
\frac {d{\bp}}{dt}={\cas{1}{2}}\int {\br}\times
\frac{\p\bom (t,{\br})}{\p t}\,\dd V,
\eqno{(1.2)}
$$
where the evolution equation for disturbed vorticity
in a steady-state external velocity field
(${\bU}$) is obtained by applying the curl operator
to the Navier-Stokes equation and by a subsequent substraction
of the equation for undisturbed flow:
$$
\frac{\p \bom}{\p t}+
({\bU}\cdot {\bnabla})\,\bom-
(\bom \cdot {\bnabla})\,{\bU}-
({\bOm} \cdot {\bnabla})\,{\bu}
+{\underline{
({\bu}\cdot {\bnabla})\,\bom-(\bom \cdot {\bnabla})\,{\bu}}
=\nu\,\Delta\,\bom}.
\eqno{(1.3a)}
$$
Here ${\bOm}={\rm curl}\,{\bU}$, and ${\bu}$
designates the disturbance-induced velocity field,
$$
\bom={\rm curl}\,{\bu}.
\eqno{(1.3b)}
$$
Because the fluid impulse is invariant
with respect to the self-induced motion of the vortex disturbance
(Batchelor  1967), the nonlinear terms (underlined in
equation (1.3a)) make a zero contribution to (1.2). This makes it
possible to ``linearize" the problem on evolution of a strongly
nonlinear localized disturbance (for a more detailed description
see the works by Levinski 1991 and Levinski \& Cohen 1995,
hereinafter L\&LC).
     The fluid impulse integral has also an additional important property.
Just by its definition (1.2), the fluid impulse describes
both the increase in amplitude and the geometrical growth of the vortex.
This is a highly important factor because
the experimentally observed growth of hairpin vortices
is not necessarily associated with the increase in vorticity amplitude.
To cover such a scenario  of a
localized vortex evolution  in terms of classical linear
stability theory requires a very extensive analysis of
amplitude changes for a great number of modes.

     The most important result, obtained on the basis of
the fluid impulse approach, is the prediction  of an exponential
instability of a localized vortex disturbance in plane Couette
flow (L\&LC). This finding makes it possible to explain
the experimentally observed formation and fast development of
hairpin vortices in the boundary layer as resulting from a plane
shear flow instability to localized vortices originating on the
wall inhomogeneities. By applying this approach to circular Cuette
flow (Malkiel, Levinski \& Cohen 1999; Levinski \& Cohen 1995), it
was possible to predict the growth of hairpin vortices within the
range of basic flow parameters where the flow is known to be
linearly stable. The criterion obtained for hairpin vortices
growth was supported by the results of experiments performed by
Malkiel, Levinski \& Cohen (1999).

     On the other hand, as the procedure
of deriving the closed evolution equation for fluid impulse,
suggested in L\&LC does not contain any formal limitations on the
initial disturbance amplitude, this result {\it is in conflict}
with the results of classical linear stability theory for
plane and circular Cuette flows
(Drazin \& Reid 1981, Dikii 1976).

     The objective of this paper is to analyze the
evolution of a localized vortex disturbance in terms of
linear stability theory on the basis of constructing
a {\it complete vorticity field}.
This approach is free from the deficiencies of the description
of vortex evolution using the fluid impulse; unfortunately,
however, it does not permit us to advance into the region
of strong (nonlinear) vortices by analytical methods.
On the other hand, the approach that is developed in this paper,
is useful for analyzing the validity of the assumptions made in L\&LC
in the course of deriving  the closed evolution equation
for fluid impulse of the disturbance. Moreover, a knowledge of a
complete vorticity field in the physical space for an arbitrary
instant of time provides the basis for constructing the other
integral characteristics of the disturbance vorticity
field which can be useful in the analysis of the results of numerical
simulations.

     This paper is organized as follows.

     In \S\,2 we subject to a critical analysis the fluid impulse
concept suggested in L\&LC, and on the basis of the exact
solution of a linearized problem
(for the time being, as a Fourier-representation)
we demonstrate its invalidity {\it for the
present formulation of the problem} (i.e. the problem of
localized vortex
evolution in the external flow with a {\it linear} velocity profile).

     In \S\,3, on the basis of  the exact solution obtained above
(in \S\,2), inverse Fourier-transform is used to construct
the vorticity field in the physical space (within
the linear approximation, of course). By considering an example where a
so-called ``Gaussian vortex" serves as the initial
disturbance, the linear evolution of the vortex is studied for
some particular cases of its  orientation.
It will be shown, in particular, that the symmetry properties of the basic
equations forbid the formation of ``hairpins" within the framework
of a linear problem.

     In \S\,4, we introduce the notion of total enstrophy
of the vortex as the measure of its intensity.
On this basis, we investigate  the character of enhancement (attenuation)
of the vortex depending on its initial orientation.

     In \S\,5, we turn from the description of the vortex
development based on its complete vorticity field to the
description using a new integral characteristic, namely, the
Tensor of Enstrophy Distribution (TED) which we introduce. This
integral characteristic makes it possible to describe the vortex
using only six independent parameters (and in the case of
vortices symmetrical about the plane $z=0$,  even four parameters
only).

      In \S\,6 we discuss the results obtained and the
possible further directions of research.
\section{The evolution of the fluid impulse. The exact solution of the
linear initial problem for the vorticity field in the ${\bf k}$-space}

\medskip

\subsection{
Definition of the modified fluid impulse of vorticity}

     It follows directly from the definition of
fluid impulse integral (1.1) that it exists
and is absolutely convergent only if
$$
|\bom (t,{\br})| \le \frac{A}{|{\br}|^{4+\epsilon}}, \ \
 {\rm where} \ \ \epsilon > 0.
\eqno{(2.1)}
$$

     Initially, a well-localized disturbance induces
a velocity field possessing the asymptotic behavior (Batchelor 1967)
$$
|{\bu}(t=0;{\br})|\sim \frac{1}{|{\br}|^3}.
\eqno{(2.2)}
$$
From substitution of (2.2) into (1.3{\it a}) it
follows that the vortex field
generated at an arbitrary time $t>0$,
has the asymptotic representation
$$
|\bom(t,{\br})|\sim \frac{1}{|{\br}|^4},
\eqno{(2.3)}
$$
which does not satisfy the condition (2.1).

     To overcome this problem the approach based on
vorticity separation procedure has been  suggested
by Levinski (1991).
Accordingly, the vorticity field of the disturbance
is subdivided on  a closed vorticity field bounded
the region
directly adjacent to the initial vortex disturbance ($\bom^I$),
and  vorticity field ($\bom^{II}$) that includes vortex
tails generated  in the
process of the vortex disturbance development.
Furthermore, it is assumed that $\bom^I$ describes the evolution of
the hairpin vortex, whereas $\bom^{II}$ represents a vortex cloud,
which has no substantial impact on the evolution of
the concentrated vorticity.

     Since the subsequent analysis is based on the exact
solution of linearized equations of vorticity dynamics and does
not involve any additional assumptions, it is appropriate to
introduce the concept of fluid impulse without recourse to the
vorticity subdivision procedure. This will permit us, in
particular, to analyze the validity of the procedure sussested by L\&LC.

For this purpose it is convenient to define
the Modified Fluid Impulse (MFI), as follows
$$
{\tilde{\bp}}(t)=\lim\limits_{R\to\infty}
{\cas{1}{2}}\int\limits_{r<R}{\br} \times\bom(t,{\br})\,\dd V.
\eqno{(2.4)}
$$

The definition (2.4) is valid both in the case of the initial
disturbance with an infinitely small amplitude
as well as in a strongly
nonlinear case. The only limitation is the local character of
the disturbance at the initial instant of time, which corresponds
to the absolute convergence  of the fluid impulse integral for the initial
distribution of vorticity.

     Since in the subsequent discussion
the solution to the equations for vorticity dynamics
is constructed in the Fourier-space,
we shall use a Fourier-transform of the definition
of the MFI. Namely,  defining the
Fourier-transform of the vorticity field as
$$
\bom(t,{\bk})=(2\upi)^{-3}\int \dd V\,\bom(t,{\br})\,
\exp\,(-{\rm i}{\bk\br}).
\eqno{(2.5)}
$$
the MFI (2.4)  be represented as
$$
{\tilde{\bp}}(t)=
{\case{1}{2}}\,\ir\,(2\pi)^3\, \lim\limits_{k\to 0}\bigl\langle
\bnabla_{k}\times\bom(t,{\bk}) \bigr\rangle,
\eqno{(2.6)}
$$
where the angle brackets correspond to averaging over the angles
in the {\bf k}-space:
$$
\langle\cdots\rangle=\frac{1}{4\upi}
\int_{-\upi/2}^{\upi/2}\cos{\beta}\,\dd\beta
\int_0^{2\upi}\!\dd\phi\,(\cdots),
\eqno{(2.7)}
$$
and
$\beta$ and $\phi$ are the spherical angles in the {\bf k}-space
(with the axis $y$ as the vertical axis, and the plane ($xz$)
corresponding to $\beta=0$):
$$
k_1=k\cos\beta\cos\phi,\ \
k_2=k\sin\beta,\ \ k_3=k\cos\beta\sin\phi.
\eqno{(2.8)}
$$

It can be shown that the
MFI defined by (2.4) is sufficiently ``good" at first glance.

     Firstly, it does satisfy the necessary requirement
of invariance with respect to the position
of the center of a sphere.

     It should be noted, however, that in the present case
where the vorticity decreases toward the periphery only
as $\sim |{\br}|^{-4}$ (and the fluid impulse
in the usual sense does not exist), the velocity field
${\bu}({\br})$ at large distances {\it is no longer a potential
one} and, in particular,
it no longer may be represented in the usual ``dipole" form
$$
{\bu}=\frac{1}{4\upi}\,{\rm curl}\,
\left(\frac{{\bp}\times{\br}}{r^3}\right)+O(1/r^4),
$$
even if the MFI ${\tilde{\bp}}$ serves as ${\bp}$.

     This means in particular that the MFI does not reflect
at all the dipole structure of the disturbance vorticity field.
As will be shown below, the plane
of the vortex core localization, which can be described
by the enstrophy distribution, $L\equiv |\bom({\br})|^2$,
is not necessarily perpendicular to the MFI direction,
as should be in the case of  the usual vortex dipole.
(It can be shown, however, that even in this case the
velocity again decreases toward the periphery
in inverse proportion to the distance cubed,
${\bu}\sim |{\br}|^{-3}$,
as in the case of the potential velocity field
induced by well-localized vortex).

     Secondly, it can be shown that the integral (2.4)
exists at any instant of time and its value
for sufficiently large values of $R$ is independent on
the value of $R$ provided the MFI is well-defined at the initial
instant of time.

     Indeed, by taking the time derivative of the
expression (2.4)  and substituting (1.3)
into the right-hand side, we obtain
$$
{\dfrac{\dd{\tilde{\bp}}(t)}{\dd t}}=
-{\cas{1}{2}}\lim\limits_{R\to\infty} \int\limits_{r<R}{\br}
\times\Bigl[({\bU}\bnabla)\,\bom-(\bom\bnabla)\,{\bU}
-({\bOm}\bnabla)\,{\bu}\Bigr]\,\dd V,
\eqno{(2.9)}
$$
Note that in (2.9) there are no contributions
from the nonlinear and viscous terms (cf. (1.3)).
The volumetric integral of these terms can be transformed
to the integral over an infinite surface.
The latter is zero by virtue of the asymptotic behavior
of the vorticity $\bom({\br},t)$ and, accordingly, of the velocity
field ${\bu}({\br},t)$ induced by it when $R\to \infty$.

Finally, the equation (2.9) may be transformed into form

$$
{\dfrac{\dd{\tilde{p}_i}(t)}{\dd t}}=
-{\cas{1}{2}}\,{\tilde p}_j\,{\dfrac{\p U_i}{\p x_j}}-
{\cas{1}{2}}\,{\tilde p}_j\,{\dfrac{\p U_j}{\p x_i}}+
\lim\limits_{R\to\infty} J_i,
\eqno{(2.10)}
$$
where
$$
J_i\!=\!-{\frac{1}{2R}}\,\ve_{ijk}{\dfrac{\p U_l}{\p x_m}}
\oint_S\! x_l x_j x_m \omega_k \dd S
$$
$$
+
{\frac{1}{4R}}\,\ve_{ijk} {\dfrac{\p U_l}{\p x_k}}
\oint_S\! x_l x_j x_m \omega_m \dd S\!+\!
{\frac{1}{4R}}\,\ve_{ijk}
{\dfrac{\p U_j}{\p x_m}}\oint_S\! x_l x_k x_m \omega_l \dd S.
$$

Here $\ve_{ijk}$ is the alternating tensor and
usual summation convention is applied.
By virtue of the asymptotic vorticity behavior, the
limit of $\bJ$ {\it is finite}.
This, together with (2.10), proves that if the MFI
exists at the initial instant of time, then it exists also at an
arbitrary instant of time.

     Thus the MFI which we have just introduced seems,
at first glance, a worthy replacement of the ``fluid impulse of the
core" introduced in L\&LC for describing
the evolution of a localized vortex since it does
not require the usage of the vortex field subdivision
procedure which has not been adequately justified in L\&LC.

     It will be shown below, however,
that {\it any} modification of the fluid impulse is
unsatisfactory with regards to the capability
to describe the structure of localized vortex.

     The  left side of equation (2.10)
together with the first two terms in its r.h.s.
represents  the evolution equation obtained in L\&LC for the
fluid impulse components constructed on the basis of a closed
vorticity field
$\bom^{I}$, whereas
the last term describes the specific contribution from
the vortex tails.

Note that the basic equation in
L\&LC is a {\it linear} equation in spite of the fact that it was
derived from the exact nonlinear system of equations (1.3)
{\it without recourse to the linearization procedure}.
This means that the theory suggested in L\&LC claims, in fact, a
possibility of describing not only weak but also strong
(nonlinear) vortices. In other words, this theory is
{\it insensitive} to the vortex disturbance amplitude.

     On the one hand, this makes it extremely attractive,
which, as a matter of fact, gave impetus to conduct a number of
elegant experiments on its basis, and, on the other, if it is
true, its predictions must remain valid for weak vortices as well.
For weak  vortices, however, there is a possibility of
drastically simplifying the problem by performing a preliminary
linearization of the initial system. This permits us to write
the {\it exact} solution for the vorticity field and, on its basis, to
check the validity of the theory. In particular, it is such a
possibility of verifying the theory suggested in
L\&LC has stimulated this investigation.

     Thus, the contribution of vortex tails
into (2.10) means that it is {\it impossible} to construct the
closed equation describing the fluid impulse evolution (and from
which the conclusion was drawn in L\&LC about exponential
instability) without one or another of
vorticity subdivision methods.

     For that reason, below we make an attempt
to give an alternative description to the evolution
of a localized vortex without recourse to the evolution
equation for fluid impulse.
Using the solution obtained for the vorticity components
we will
also be able to calculate the fluid impulse
and check to what extent the solution describing
the fluid impulse evolution obtained in L\&LC
is consistent with what follows from this solution.
%
\subsection{The evolution
of a localized vortex disturbance in the plane Couette flow}
%
\subsubsection{ Solving the initial problem for a localized
disturbance}

     It will be assumed that the
basic flow has a linear velocity profile, ${\bU}=(-\Omega\,y,0,0)$
(so that its vorticity is ${\bOm}=(0,0,\Omega)$).
Note that this choice is not a loss of
generality as a consequence of the assumption about the local
character of the disturbance. It corresponds to the case where
the characteristic size of the disturbance is much less than the
characteristic size of variation of the basic velocity field.

The suitable mathematical method
for investigation of disturabance evolution in  such flow
was proposed by Lord Kelvin more
than centure ago
(Kelvin 1887,
see also the more recent publications based on this method,
such as Craik \& Criminale 1986, Criminale \& Drazin 1990,
Farrel \& Ioannou 1993
and other), and we also apply it here.

In
this case the inviscid linearized equation for vorticity dynamics
can be represented in the component-wise notation in Cartesian
coordinates as
$$
\left.
\ba{c}
\dfrac{\p \omega_1}
{\p t}-y\,\Omega\,\dfrac{\p \omega_1}{\p x} -\Omega\,
\dfrac{\p u_3}{\p x}=0,\phantom{\Bigg|}\\
\dfrac{\p \omega_2}{\p t}-y\,\Omega\,\dfrac{\p \omega_2}{\p x}
-\Omega\,\dfrac{\p u_2}{\p z}=0,\phantom{\Bigg|}\\
\dfrac{\p \omega_3}{\p t}-y\,\Omega\,\dfrac{\p \omega_3}{\p x}
-\Omega\,\dfrac{\p u_3}{\p z}=0,\phantom{\Bigg|}
\ea
\right\}
\eqno{(2.11)}
$$
where
the subscripts ``1", ``2" and ``3" correspond to the $x$-,
$y$- and $z$-components of the vectors, respectively. Upon
introducing the dimensionless time $\tau=\Omega t$ and
Fourier-transforming the system of equations (2.12), we
obtain
$$
\left.
\ba{c}
\dfrac{\p\omega_1}{\p\tau}+
k_1 \dfrac{\p \omega_1}{\p k_2}-\ir\,k_1 u_3=0,\phantom{\Bigg|}\\
\dfrac{\p\omega_2}{\p\tau}+k_1\dfrac{\p \omega_2}{\p k_2}-\ir\,k_3
u_2=0,\phantom{\Bigg|}\\
\dfrac{\p\omega_3}{\p\tau}+k_1\dfrac{\p \omega_3}{\p k_2}-\ir\,k_3 u_3=0,
\phantom{\Bigg|}
\ea
\right\}
\eqno{(2.12)}
$$
where $u_i(t,{\bk})$ represent the Fourier-transform
of the disturbed velocity field components $u_i({\br},t)$.

     The solution of the above problem on the evolution
of the perturbation in the form of a single plane wave
has been presented earlier in the work by
Farrel \& Ioannou, 1993 (hereinafter F\&I). Although the solution
presented here below, in fact, is the same, we describe it derivation
briefly to emphasize here on
the {\it vorticity}  components (instead of
velocity components, as it done in F\&I).
We  use, as in F\&I,
instead of the set of independent
variables $\tau$, $k_1$, $k_2$ and $k_3$,
a new set of independent variables
$\tau$, $k_1$, $q$ and $k_3$, where $q=k_2-k_1\tau$.
We will consider $\omega_i$ as a function
of $\tau$, $k_1$, $q$, $k_3$, that is,
$\omega_i=\omega_i(\tau,k_1,q,k_3)$.
Thus the new variable $q$ is simply the
initial value of the time-varying component of the wave vector $k_2$:
$k_2(t)=q+k_1\tau$.

     The new set of variables will be referred to as
the Lagrangian coordinates in the {\bf k}-space.
The transition to the Lagrangian coordinates makes
it possible to transform the system of equations (2.12)
to a system of ordinary differential equations
$$
\left.
\ba{l}
\dfrac{\dd\omega_1}{\dd\tau}-\dfrac{k_1}{k^2(\tau)}\,
\bigl[k_2(\tau)\,\omega_1-k_1\omega_2\bigr]=0,\phantom{\Bigg|}\\
\dfrac{\dd\omega_2}{\dd\tau}-\dfrac{k_2(\tau)}{k^2(\tau)}\,
\bigl[k_2(\tau)\,\omega_1-k_1\omega_2\bigr]+\omega_1=0,\phantom{\Bigg|}\\
\dfrac{\dd\omega_3}{\dd\tau}-\dfrac{k_3}{k^2(\tau)}\,
\bigl[k_2(\tau)\,\omega_1-k_1\omega_2\bigr]=0,\phantom{\Bigg|}
\ea
\right\}
\eqno{(2.13)}
$$
where
$
k^2(\tau)=k_1^2+[k_2(\tau)]^2+k_3^2,
$
and
$
\dd/\dd\tau\equiv (\p/\p\tau)_q=(\p/\p\tau)_{k_2}+k_1(\p/\p k_2)_{\tau}
$.
    In  deriving (2.13), we expressed also the
Fourier-components of the velocity field in terms of
Fourier-components of the vorticity field
$ {\bu}({\bk})=
\ir\,\dfrac{{\bk}\times\bom({\bk})}{k^2}.
$
As a result we obtain
the solution for
dynamics of vorticity $\bom(\tau,{\bk})$ as:
$$
\left.
\ba{c}
\omega_1(\tau,{\bk})=\omega_1({0,\bf Q})-
\dfrac{k_1^2}{p^2}\ \tau\,\omega_2({0,\bQ})+
\dfrac{k_2k_3}{p}\,V(\tau,{\bk})\,T(\tau,{\bk}),\phantom{\Bigg|}\\
\omega_2(\tau,{\bk})=\omega_2(0,{\bQ})-\dfrac{k_3 p}{k_1}\,
V(\tau,{\bk})\,T(\tau,{\bk}),\phantom{\Bigg|}\phantom{\Bigg|}\\
\omega_3(\tau,{\bk})=\omega_3(0,{\bQ})-
\dfrac{k_1 k_3}{p^2}\,\tau\,\omega_2(0,{\bQ})+
\dfrac{k_2 k_3^2}{k_1 p}\,V(\tau,{\bk})\,T(\tau,{\bk}).
\ea
\right\}
\eqno{(2.14)}
$$
Here
$p=\sqrt{k_1^2+k_3^2}$, ${\bQ}=(q_1,q_2,q_3)\equiv (k_1,q,k_3)$
is the initial value of the wave vector which in (2.14)
must be expressed in terms of ${\bk}$:
${\bQ}=(k_1,k_2-k_1\tau,k_3)$,
$$
V(\tau,{\bk})\!=\!(1/p^2)\bigl[k_3\omega_1(0,{\bQ})\!-\!
k_1\omega_3(0,{\bQ})\bigr],\ \
T(\tau,{\bk})\!=\!\arctan(k_2/p)\!-\!\arctan\bigl[(k_2\!-\!k_1\tau)/p\,\bigr].
$$
Including of the viscosity leads to
addidional viscous factor in the expressions for
$\omega_i$:
$$
\omega_i(\tau;\bk)\to \omega_i(\tau;\bk)\exp\Bigl(-(\nu/\Omega)\int_0^{\tau}
k^2(\tau')d\tau'\Bigr)
$$
(see also F\&I).

In what follows, we shall use, as the initial vortex
disturbance, the Gaussian vortex
$$
\bom(\tau=0,{\br})=\bnabla F\times \bmu,\ \
F=(\upi^{1/2}\delta)^{-3}\exp\,(-r^2/\delta^2).
\eqno{(2.15)}
$$

     For numerical simulation of all hydrodynamic
quantities (using program packages for 3-D hydrodynamics)
it is often also necessary to specify the initial velocity field.
It is readily calculated even for the initial isotropic
function $F(r)$:
$$
{\bu}=F(r)\Bigl[\,\bmu-\frac{{\br}\,(\bmu{\br})}{r^2}\Bigr]-
\frac{H(r)}{r^3}\,\Bigl[\,\bmu-\frac{3{\br}\,(\bmu{\br})}{r^2}\Bigr],
\ \
H(r)={\int_0^r} F(x)\,x^2\,\dd x
\eqno{(2.16)}
$$

     Thus we have in the ${\bk}$-representation
for the vortex of (2.15)
$$
\bom(0,{\bk})
=\frac{\ir}{(2\upi)^3}\, ({\bk}\times\bmu)
\exp\,(-\cas{1}{4}k^2\delta^2).
\eqno{(2.17)}
$$
Note that for the vortex (2.15) at the initial instant
of time
the {\it usual} fluid impulse
$\bp$ is also well defined and equal to $\bmu$.

     For our further purposes it is also very
convenient to use the spherical coordinates in the ${\bQ}$-space:
$
q_1(=k_1)=Q\cos\beta_0\cos\phi,\ \ q_2=Q\sin\beta_0,\ \
q_3(=k_3)=Q\cos\beta_0\sin\phi,
$
and $-{\cas{1}{2}}\upi\le\beta_0\le{\cas{1}{2}}\upi$,
$0\le\phi\le 2\upi$.
     In these variables we write finally for the Gaussian
vortex
$$
\omega_i({\tau,\bk})=\frac{\ir}{(2\upi)^3}\,p\,
\exp\,(-\cas{1}{4}Q^2 D^2)\, {\hat\zeta}_i(\tau;\beta_0,\phi),
\eqno{(2.18)}
$$
$$
\left.
\ba{c}
{\hat\zeta}_1=\mu_3\tan\beta_0-\mu_2\sin\phi-
\tau\cos^2\!\phi\,(\mu_1\sin\phi-\mu_3\cos\phi)-
\tan\beta\cos\phi\,\zeta_2^{\ast},\phantom{\bigg |}\\
{\hat\zeta}_2=\mu_1\sin\phi-\mu_3\cos\phi+\zeta_2^{\ast},\phantom{\bigg |}\\
{\hat\zeta}_3=\mu_2\cos\phi-\mu_1\tan\beta_0-
\tau\sin\phi\cos\phi\,(\mu_1\sin\phi-\mu_3\cos\phi)
-\tan\beta\sin\phi\,\zeta_2^{\ast},\phantom{\Bigg |}\\
\zeta_2^{\ast}=(\beta_0-\beta)\,
\tan\phi\,\bigl[\tan\beta_0\,(\mu_1\cos\phi+\mu_3\sin\phi)-\mu_2\bigr].
\ea
\right\}
\eqno{(2.19)}
$$
Here
$$
D=D(\tau;\beta_0,\phi)=\delta\,\Bigl\{1+\dfrac{4\tau}{Re}
\,\bigl[1\!+\!\tau\cos\beta_0\sin\beta_0\cos\phi
\!+{\cas{1}{3}}\tau^2\cos^2\!\beta_0\cos^2\!\phi\bigr]\Bigr\}^{1/2},
\eqno{(2.20)}
$$
and the ``Reynolds number of vortex"  $Re$ is defined as
$Re={\Omega\,\delta^2}/{\nu}$ (so that in the inviscid case $D=\delta$).
The spherical angles $\beta$ (in the ${\bk}$-space)
and $\beta_0$ (in the ${\bQ}$-space)
are related by the
following expression
$
\beta=\beta\,(\tau;\beta_0,\phi)=\arctan\,(\tan\beta_0+\tau\cos\phi).
$
\subsubsection{Evolution of the
Modified Fluid Impulse of a localized disturbance}

     As has already been discussed in the \S\,1,
the objective of this investigation was, in particular, to
calculate the dynamics of fluid impulse of localized vortex in the
external shear flow {\it without using the vorticity subdivision
procedure} employed in L\&LC.
Indeed, the general solution of the  vorticity dynamics
equations, presented in \S\,2.2.1,
makes it possible to calculate the MFI defined by (2.4)
at an arbitrary instant of time.

     For illustrative purposes we avail ourselves
of the Gaussian vortex model (2.15) introduced above.
It should be noted that the theory by L\&LC
is insensitive not only to the vortex disturbance amplitude
but also to its form. Therefore, the initial disturbance
can be chosen rather arbitrarily. Choosing it in the form a
Gaussian vortex (2.15) optimizes calculations substantially,
but from the other side it is sufficiently representative model
(see also notation in \S\,6).
The initial vortex is shown in figure 1,
portraying the enstrophy isosurface
$\bom^2(0,{\br})={\rm const}$
that represents the surface of a torus.
\begin{figure}
\epsfysize=35mm
\centerline{\epsfbox{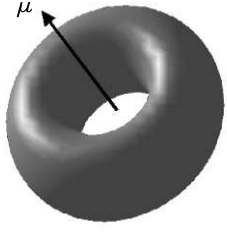}}
\caption{
Surface of constant enstrophy
$|\omega(0,{\bf r})|^2={\rm const}$
of the initial Gaussian vortex.}
\label{Fig. 1}
\end{figure}
Note that for this model the size of the vortex disturbance core
is specified by the value of the parameter $\delta$, and the
vortex plane and vortex lines that represent concentric circles
are normal to the direction of the initial fluid impulse $\bmu$.

     Substitution of the expressions (2.18)--(2.19)
for vorticity components into the expression (2.6) for the MFI gives:
$$
{\tilde p}_i(\tau)=\sum\limits_{j=1}^{3}{\tilde\Pi}_{ij}(\tau)\,\mu_j,
\eqno{(2.21)}
$$
$$
\left.
\ba{c}
{\tilde\Pi}_{11}=1+
{\cas{3}{2}}\,\bigl\langle\sin^2\!\phi\tan\beta_0\,(\beta_0-\beta)
\bigr\rangle=1+\cas{1}{2}I_1,\phantom{\bigg|}\\
{\tilde\Pi}_{12}={\cas{3}{2}}\,\Bigl\langle{\sin\phi}\,
{\tan\phi}\,(\beta-\beta_0)\Bigr\rangle={\cas{1}{2}}I_2,\phantom{\bigg|}\\
{\tilde\Pi}_{21}=\cas{1}{2}\tau,\ \ {\tilde\Pi}_{22}=1,\ \
{\tilde\Pi}_{33}=2-{\tilde\Pi}_{11}=1-\cas{1}{2}I_1
\ea
\right\},
\eqno{(2.22)}
$$
where the angle brackets correspond to averaging
over angles in the ${\bk}$-space,  and
$
\beta_0=\beta_0\,(\tau;\beta,\phi)=\arctan\,(\tan\beta-\tau\cos\phi).
$
In the expanded form we have
$$
{\tilde p}_1(t)=\mu_1+\cas{1}{2}\mu_1\,I_1(t)+
\cas{1}{2}\mu_2\,I_2(t),\ \
{\tilde p}_2(t)=\mu_2+\cas{1}{2}\mu_1\,\Omega t,\ \
{\tilde p}_3(t)=\mu_3-\cas{1}{2}\mu_3\,I_1(t).
\eqno{(2.23)}
$$
     It is easy to obtain the asymptotic
expressions for $I_1$ and $I_2$ for small
($t\ll 1/|\Omega|$) and large ($t\gg 1/|\Omega|$) times:
$$
I_1(t)=\frac{1}{5}\,(\Omega t)^2+
{\cal O}\bigl((\Omega t)^3\bigr),\ \
I_2(t)=\Omega t+
{\cal O}\bigl((\Omega t)^3\bigr),\ \ |\Omega|t\ll 1,
\eqno{(2.24)}
$$
and
$$
I_1(t)=|\Omega|t-3+O(1/|\Omega|t),\ \
I_2(\tau)=3\bigl[\ln(|\Omega|t)-0.6\bigr]+O(1/|\Omega|t),
\ \ |\Omega|t\gg 1.
\eqno{(2.25)}
$$

     For small $t$ we have from (2.23) and (2.24):
$$
{\tilde p}_1\!\approx\!\mu_1\!+\!\cas{1}{2}\mu_2 (\Omega t),\
{\tilde p}_2\!\approx\!\mu_2+\cas{1}{2}\mu_1 (\Omega t),\
{\tilde p}_3\!=\!\mu_3.
\eqno{(2.26)}
$$
     As would be expected, the MFI dynamics
in the case of small times is determined by the first two terms
in the evolution equation (2.10), in full agreement with the
theory by L\&LC. This is because the initial vortex (2.15)
is well localized; therefore, the contribution
$\bJ$ associated with the presence (at the
early stage of evolution still very weak) vorticity ``tails" at
the periphery of the vortex, $\bom\sim r^{-4}$, is vanishingly
small.

     In the case of larger times, however,
the situation changes radically. The vorticity ``tails"
now become quite important and begin to affect
the fluid impulse dynamics.

     This leads, in particular, to the fact that,
according to (2.25), at large $t$ the fluid impulse
increases {\it as a power law} rather than exponentially,
as was the case in L\&LC.

     The fact that both the vorticity itself
(including, of course, its ``tails" produced in the course of evolution)
and the MFI increase not more rapidly than as a power law,
makes it possible to prove rigorously
that there is no way to subdivide the disturbance vorticity field
into two different components as it was suggested in L\&LC. Namely,
it is impossible to separate a localized vortex core, the fluid impulse
of which grows exponentially, from the complete vorticity
field (see \S\,6 for more details).

     But it is the {\it assumption}
about the possibility of such a separation that formed
the basis of the approach suggested in L\&LC.

     This means that the approach by L\&LC
is incorrect, in spite of a number of
predictions obtained on this base which show an excellent
agreements with
experimental findings. Consequently, the problem of
constructing an adequate theory describing the dynamics of
localized vortices in shear flows becomes of current importance
again.

     Nevertheless, it is interesting to point out
that the inclination angle $\bPsi$ (of the MFI vector
to the positive direction of the $x$-axis tends to
$45^{\circ}$  with the time, in exactly the same way
as does the fluid impulse ${\bp}^I$ constructed from the ``core
vorticity"  in L\&LC.

     Indeed, when $|\Omega|t\gg 1$ from (2.38) using (2.40) we have
$$
\left.
\ba{c}
{\tilde p}_1(t)\!=\!-\cas{1}{2}\mu_1+\cas{1}{2}\mu_1\,|\Omega| t
\!+\!\cas{3}{2}\mu_2\,\bigl[\ln(|\Omega|t)-0.6\bigr],\ \
{\tilde p}_2(t)\!=\!\mu_2+\cas{1}{2}\mu_1\Omega t,{\phantom{\Big|}}\\
{\tilde p}_3(t)\!=\!\cas{5}{2}\mu_3\!-\!\cas{1}{2}\mu_3\,|\Omega|t,
\ea
\right\}
\eqno{(2.27)}
$$
so that for vortices symmetric about the plane $z=0$ ($\mu_3=0$)
we obtain:
$
\tan\bPsi={\tilde p}_2(t)/{\tilde p}_1(t)\to 1.
$
     Figure 2 shows the evolution of the
quantities $I_1(\tau)$ and $I_2(\tau)$
as well as of the inclination angle $\bPsi$
for the case $\bmu=(1,0,0)$.
\begin{figure}[t]
\epsfxsize=50mm
\centerline{\epsfbox{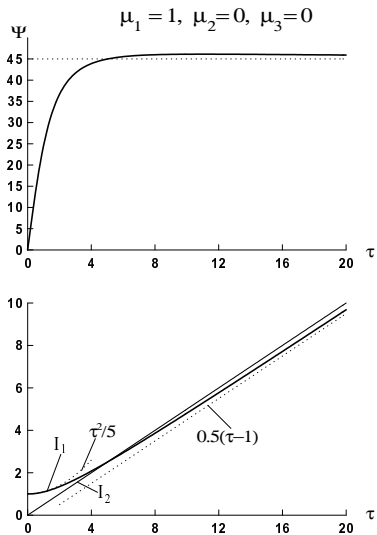}}
\caption{
The evolution of $I_1(\tau)$ and $I_2(\tau)$ and of the
inclination angle $\Psi$ of the modified fluid impulse
for the vertically oriented ($\mu_1=1,\ \mu_2=0$)
symmetric ($\mu_3=0$) Gaussian vortex.}
\label{Fig. 2}
\end{figure}

     However, in spite of the relatively good agreement
between the orientation of the vortex plane observed
in experiments and its orientation following from the
description of the vortex using the MFI,
it {\it does not mean} at all that the MFI is a
more acceptable characteristic for describing
localized vortices which should replace the ``unfortunate"
description using the ``fluid impulse of the core"
${\bp}^{I}$ suggested previously in L\&LC.

     It will be shown below that it is also
possible to suggest some other methods of modifying the fluid
impulse which will also be free from difficulties associated with
the convergence of the corresponding integral at large distances,
as is the just considered MFI; however, they will lead to a
totally different scenario for the vortex geometry evolution.

     Actually, this would mean that the fluid impulse
is not an adequate characteristic at all for describing the
evolution of localized vortices. At least for the statement of
the problem of localized vortex
disturbance development in
the external flow with the {\it linear}
velocity profile accepted here (as well as in previous
publications  L\&LC, Levinski, Rapoport \& Cohen 1995,
Malkiel, Levinski \& Cohen 1999 and Levinski 2000).
\subsubsection{The evolution of the
``Lagrangian" Modified Fluid Impulse of a localized disturbance}
     In this section we introduce
the concept of the fluid impulse of a selected (``colored")
group of fluid particles and investigate its evolution.

     Let the initial position of a fluid particle
that resides at the time $\tau$ at a point with
the coordinates ${\br}=(x_1,x_2,x_3)$ be
designated as
${\br}_0$: ${\br}\,(\tau=0)\equiv {\br}_0$,
where  the components of the vector ${\br}_0$
be $s_1$, $s_2$ and $s_3$: ${\br}_0=(s_1,s_2,s_3)$. Then
$
x_1=s_1-s_2\tau,\ \ x_2=s_2,\ \ x_3=s_3,
$
     We now select a group of particles which
at the initial instant of time are enclosed
within a sphere of radius $R$ and mentally paint it:
$
r_0\equiv\sqrt{s_1^2+s_2^2+s_3^2}\le R.
$
     We shall keep track on this painted group
and calculate its fluid impulse. At subsequent instants
of time, when $\tau\ne 0$, the painted sphere will
transform to an ellipsoid
$(x_1+x_2\tau)^2+x_2^2+x_3^2\le R^2$.

     Thus the fluid impulse of the painted group of particles is
$$
{\hat p}_i(\tau,R)={\cas{1}{2}}\,\epsilon_{ijk}
\int_{ell}x_j\omega_k(\tau;{\bf r})\,\dd V,
$$
where the integral is taken over the volume of the
ellipsoid. By letting further $R\to\infty$
%
we obtaine for the LMFI components:
$
{\hat p}_i=
{\hat \Pi}_{ij}\mu_j.
$
Omitting  the explicit expressions for ${\hat\Pi}_{ik}$,
we present here only their
asymptotic expressions for $\tau\gg 1$:
$$
\left.
\begin{array}{c}
{\hat\Pi}_{11}\approx-{\cas{1}{6}}\,\tau+2,\ \
{\hat\Pi}_{12}\approx-{\cas{1}{6}}\,\tau+{\cas{3}{2}}\ln\tau-0.414,\ \
{\hat\Pi}_{21}\!\approx\!-{\cas{1}{6}}\,\tau^2\!+\!{\cas{4}{3}}\,\tau
\!-\!{\cas{3}{4}}\ln\tau,
{\phantom{\Bigg|}}\\
{\hat\Pi}_{22}\!\approx\!-{\cas{1}{6}}\,\tau^2\!+
\!{\cas{1}{2}}\,\tau\ln\tau
\!+\!0.026\,\tau,\ \
{\hat\Pi}_{33}\!\approx\!-{\cas{1}{2}}\,\tau\ln\tau\!+\!0.64\,\tau.
{\phantom{\Big{|}}}
\ea
\!\!\!\!\right\}
\eqno{(2.28)}
$$

It is evident that for
large $\tau$
$$
{\hat p}_1\approx -{\cas{1}{6}}\tau\,(\mu_1+\mu_2),
\ \ {\hat p}_2\approx\tau {\hat p}_{1}\approx-
{\cas{1}{6}}\tau^2\,(\mu_1+\mu_2),
\ \ {\hat p}_3\approx-{\cas{1}{2}}\tau\ln\tau\,\mu_3.
$$

     We can now easily calculate the inclination
angle $\bPsi$ of the vector
${\hat{\bp}}=({\hat p}_1,{\hat p}_2,{\hat p}_3)$
in the $(xy)$-plane for
different orientation angles of the initial fluid impulse
${\hat{\bp}}(0)(\equiv\bmu)$:
$
\tan\bPsi(\tau)={\hat p}_2(\tau)/{\hat p}_1(\tau)=
[{\hat\Pi}_{21}\mu_1+{\hat\Pi}_{22}\mu_2]/
[{\hat\Pi}_{11}\mu_1+{\hat\Pi}_{12}\mu_2].
$

     Figure 3 shows the evolution of $\bPsi(\tau)$ for
8 orientations of
$\bmu$: $\alpha_n=(n-1){\cdot}45^{\circ}$.
\begin{figure}[t]
\epsfxsize=60mm
\centerline{\epsfbox{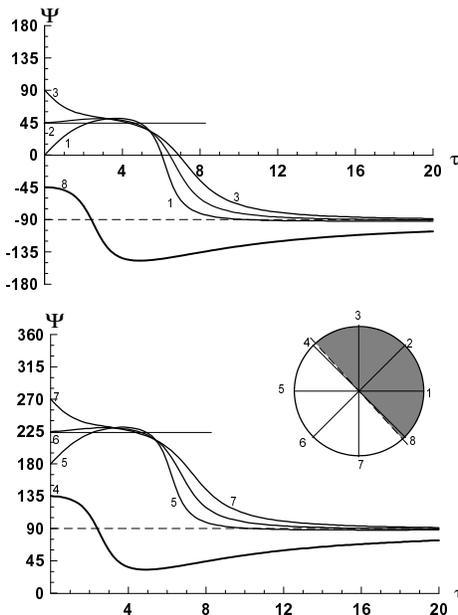}}
\caption{
The evolution of the
inclination angle $\Psi$ of the vector
${\hat{\bf p}}$. Numbers at the curves
correspond to the initial orientation.}
\label{Fig. 3}
\end{figure}
It is evident that at large times the two-dimensional
vector ${\hat{\bp}}=({\hat p}_1,{\hat p}_2)$ is directed {\it
vertically}. The vortices, the directions of the initial fluid
impulse of which lie in figure 3 in the shaded and unshaded
areas are directed at large $\tau$ downward and
upward respectively.

     It is interesting to note that over a sufficiently
long time interval $2<\tau<6$ the inclination angle is
about $45^{\circ}$ or $225^{\circ}$
(according to the initial inclination angle $\alpha$).


\smallskip

\smallskip

     {\it Summary}.
Thus it is evident that two different methods
of modifying the fluid impulse, MFI and LMFI, lead to two
totally different results as regards their orientation
at asymptotically large times:
$45^{\circ}$ for MFI, and $90^{\circ}$ for
LMFI. It will be recalled that we associated intuitively
the orientation of the vortex plane with the orientation of
the fluid impulse vector ${\bp}$ by assuming that, as for the
usual dipole structure (such as in magnetostatics if
we mean the analogy:
${\bu}\to {\bH}$, $\bom\to {\bj}$, ${\bp}\to {\bm}$,
where ${\bH}$
is the magnetic field, ${\bj}$ is electric
current density, and
 ${\bm}={\cas{1}{2}}
\int ({\br}\times{\bj})\, \dd V$
is the magnetic dipole moment),
this plane must simply be normal to the fluid impulse
direction (L\&LC) just as the plane of a ringlet with
current is perpendicular to the dipole magnetic moment.

     It now becomes clear, however, that the fluid
impulse in this problem just cannot describe adequately
the vorticity distribution. By choosing in a different
manner the form of the domain of integration, we can obtain
for the same vorticity
distribution not only an arbitrary time dependence
of its fluid impulse but also an arbitrary inclination
of the vortex plane.

     For that reason, there inevitably arises the
problem of calculating a complete vorticity field. It is a fairly
complicated numerical problem which is being solved to
date (preliminary results of these calculations are
presented  in Suponitsky
\etal, 2003, 2004), however within the linear approximation, it is
actually solved (for single plane wave) by F\&I
and also in \S\,2.2.1. It will now suffice to perform
an inverse Fourier-transform and calculate the vorticity field in
the physical space. This is done in \S\,3.
\section{Calculating the complete vorticity field in physical
space}

     We have
$$
\omega_i({\br})=\int\!\omega_i({\bk})\exp\,(\ir\,{\bf kr})\,\dd^{3}k,
\eqno{(3.1)}
$$
where
$$
\omega_i({\bk})=\frac{\ir}{(2\upi)^3}\,p\,\zeta_i({\bk}),\ \
\zeta_i({\bf k})
={\hat\zeta}_i(\beta,\phi;\tau)\exp\,(-\cas{1}{4}Q^2\delta^2),
\eqno{(3.2)}
$$
and the quantities ${\hat\zeta}_i$ are specified
by the expressions (2.19).

     The most compact method for evaluating the
integral of (3.1) lies in passing from integrating in the
${\bk}$-space,
to integrating in the $\bQ$-space (i.e. in the space of
{\it initial} wave numbers).
Introducing the spherical coordinates $r_0$, $\theta_0$ and $\varphi_0$
of the point ${\br}_0$:
$$
\left.
\ba{c}
r_0=\sqrt{(x_1+x_2\tau)^2+x_2^2+x_3^2},\ \
\cos\theta_0=\dfrac{x_2}{r_0},\ \
\\
\cos\varphi_0=\dfrac{x_1+x_2\tau}{\sqrt{(x_1+x_2\tau)^2+x_3^2}}
\ea
\right\}
\eqno{(3.3)}
$$
and also the angle $\Theta_0$  between the vectors
${\bQ}$  and ${\br}_0$:
$$
\cos\Theta_0=\cos\theta_0\sin\beta_0+\sin\theta_0\cos\beta_0
\cos(\phi-\varphi_0).
\eqno{(3.4)}
$$
we obtain
$$
\begin{array}{c}
{\omega}_i(\tau;{\br})
=-\dfrac{2}{\upi^{5/2}}{\displaystyle\int_{0}^{\upi/2}}\!\cos^2\!\beta_0\,
\dd\beta_0\\
\\
\times
{\displaystyle\int_0^{2\upi}}\dfrac{\dd\phi}{D^4}\,
{\hat\zeta}_i(\beta_0,\phi;\tau)\Bigl(\dfrac{r_0\cos\Theta_0}{D}\Bigr)
\Bigl(\dfrac{3}{2}\!-\!\dfrac{r_0^2\cos^2\!\Theta_0}{D^2}\Bigr)
\exp\Bigl(-\dfrac{r_0^2\cos^2\!\Theta_0}{D^2}\Bigr),
\end{array}
\eqno{(3.5)}
$$

     In spite of the fact that the expression (3.5)
is a sufficiently compact one, it still is very difficult for analysis,
as it includes  double integrals. For that reason, we have to
carry out the subsequent analysis numerically.

     Note that the applicability of linear theory is
limited by the condition $|\bom|_{\rm max}\ll |\Omega|$.
This means, in particular, that this condition must also be satisfied for the
initial vortex, i.e.
$$
|\bom|_{\rm max}=|\bom(r=\delta/\sqrt{2})|=
\sqrt{\frac{2}{\upi^3e}}\,\frac{\mu}{\delta^4}
=0.154\,\frac{\mu}{\delta^4}\ll\Omega,\ \
{\rm or}\ \
\mu\ll 6.49\,\Omega\,\delta^4.
$$

In order to investigate the linear evolution of the
vortex, we calculated numerically the
vorticity field distribution for fixed
instants of  time $\tau$
by formula (3.5). Results are
presented in Fig.~4 in the form of 3-D
isosurfaces of absolute value of vorticity (or,
that is the same, of the enstrophy density $L$) for fixed
instants of time $\tau$:
$
L(\tau;{\br})\equiv |\bom(\tau;{\br})|^2={\rm const}.
$
\begin{figure}
\epsfxsize=140mm
\centerline{\epsfbox{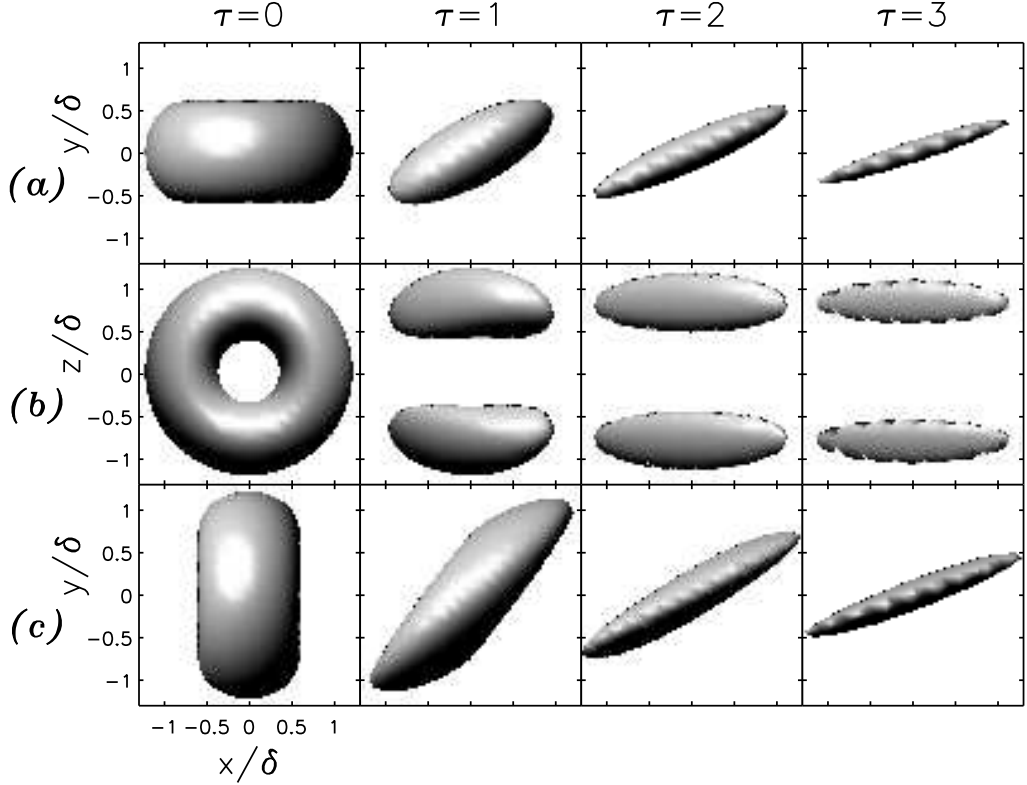}}
\caption{
The linear evolution of the Gaussian vortex:
$(a)$, $(b)$ -- {\it horizontal}
and  $(c)$ -- {\it vertical}.
Isosurfaces of absolute value of vorticity
are shown
$\omega(\tau;{\bf r})=0.7\, \omega_{\rm max}\,(\tau)={\rm const}$,
where $\omega_{\rm max}(\tau)$
is maximum (over volume)
absolute value of vorticity at time  $\tau$.}
\label{Fig. 4}
\end{figure}

     From here on we shall confine ourselves to
the case of {\it symmetric} initial vortices, $\mu_3=0$.
Furthermore, as is easy to understand, the vortex remains
symmetric about the plane $z=0$ over the course of all subsequent
evolution as well. Specifically, for the enstrophy density  we
have $L(\tau;x,y,z)=L(\tau;x,y,-z)$.

     It is apparent from figure 4 that with the passage of time,
the initial torus, corresponding to the Gaussian vortex,
starts  to rotate and deform and eventually turns into two
symmetric ``sausages" extended along the flow.
(Note that {\it only} in this figure it is assumed
that $-\Omega\equiv dU/dy > 0$
in order to achieve a more {\it usual} perception
of the vortex plane orientation.)

     These ``sausages" could, in principle,
serve as a source material for the hairpin legs.
It can be shown, however, that {\it within the framework
of linear theory} the sausages cannot turn into
a hairpin through the formation of a bridge (a so-called hairpin
head) near {\it only one of the ends} of the pair of legs.
It turns out that this is forbidden by the symmetry
properties of the basic equations!

     Indeed, it follows from the {\it linearized} set of equations (1.3)
that if
$
\bom(0;{\br})$$=-\bom(0;-{\br})
$
and
${\bu}(0;{\br})={\bu}(0;-{\br}),
$
i.e. if all components of the initial vorticity change
their sign with
${\br} \to -{\br}$
and, accordingly,  all components of the initial velocity do not alter
their values in the case of the substitution
${\br} \to -{\br}$, then this
symmetry property remains during
the vortex ({\it linear}) evolution:
$$
\bom(t;{\br})=-\bom(t;-{\br})\ \ {\rm and}\ \
{\bu}(t;{\br})={\bu} (t;-{\br}).
\eqno{(3.6)}
$$
Note that the Gaussian initial vortex does
{\it possess} the aforementioned symmetry properties (3.6).

     Consequently, for the enstrophy density
$L(t;{\br})$ of the initial
Gaussian vortex at an arbitrary instant of time $t$ we have
$
L(t;x,y,z)=L(t;-x,-y,-z).
$
    In the case of a symmetric vortex, $\mu_3=0$,
we have an additional $z$-symmetry:
$L(t;-x,-y,-z)$ $=L(t;-x,-y,z)$.
Con\-se\-q\-u\-e\-n\-t\-ly, we ob\-ta\-in
$
L(t;x,y,z)=L(t;-x,-y,z).
$
This means that a current enstrophy distribution
must be invariant with respect to the simultaneous
replacement $x\to -x$, $y\to -y$
{\it even for the same $z$}. Specifically,
directly ``between the legs", i.e. in the plane $z=0$, we have
$$
{L}(t;x,y,0)=L(t;-x,-y,0).
\eqno{(3.7)}
$$

     Hence  we {\it cannot} obtain the ``hairpin"
in the course of the Gaussian vortex evolution, since the
distribution of enstrophy in hairpin vortex
does not have the symmetry property (3.7).

     However, the situation is changed drastically
if we include in consideration the nonlinear terms
in equation (1.3){\it a}). It is easy to see that the nonlinear
terms (underlined in (1.3{\it a})) totally {\it destroy} the symmetry
properties of the linearized version of equation
(1.3{\it a}) and, hence, the prohibition for the hairpin
in linear theory is {\it removed}!

     Therefore, (numerical) investigation of
the {\it nonlinear} evolution stage of a localized vortex
is strongly needed. Preliminary results of numerical
calculations with strong vortices confirm the occurrence
of hairpins at a definite evolution stage
of the vortex (Suponitsky \etal, 2003, 2004).
%
\section{Total enstrophy of a localized vortex and its growth}
\subsection{Calculating the total enstrophy of a localized
vortex in inviscid flow}

     In order to be able to describe
the enhancement or attenuation of the vortex
over the course of the evolution, we introduce,
as one of its integral characteristics, the total enstrophy:
$$
{\cal cal L}=\int{\bom}^2({\br})\,\dd V
\eqno{(4.1)}
$$

     The total enstrophy ${\cal L}$ can serve as the
measure of vortex intensity.
Note that, at first glance,
it seems more natural to take, as the measure of vortex
intensity, its total energy ${\cal E}$. However, an attempt to
introduce its reasonable definition, like
${\cal E}=\int [({\bf U}+{\bf u})^2-{\bf U}^2]\,\dd V
{\phantom{\big|}}$,
runs into the same difficulty
into which we ran in our attempt to introduce the fluid impulse.
Indeed, since $\int {\bf (U\cdot u)}\, \dd V$ is divergent
(remember that ${\bf u}\sim r^{-3}$ for large $r$), such a
definition of the vortex energy cannot be recognized as correct.
It is also easy to see that a similar introduction of the total
enstrophy ${\cal L}$ is free from such difficulties because
$\int\!\bom\, \dd V=0$.

The total enstrophy
${\cal L}$ depends on the time $\tau$  as well as on the
the initial fluid impulse $\bmu$. If the viscosity $\nu$ is
included, then ${\cal L}$ depends
also on the Reynolds number $Re$.
It is clear that finite viscosity effects are
highly important for the problem of the maximal vortex enhancement
(cf. also with F\&I), especially in the connection
with the problem of hairpin formation in the course
of nonlinear evolution of initial weak vortex
(which is partly described by Suponitsky \etal, 2003, 2004
and will be considered in details in the following publication).
\smallskip

     Thus we have ${\cal L}={\cal L}(\tau,\bmu)$.
We can express the integral
in terms of Fourier-variables:
$$
{\cal L}=(2\upi)^3\int|{\bom}({\bk})|^2\,\dd^{\,3}k.
\eqno{(4.2)}
$$
and use the corresponding expressions for Fourier-components of vorticity.
We also introduce the {\it normalized} total enstrophy:
$$
{\hat{\cal L}}(\tau,{\bmu})={\cal L}(\tau,\bmu)/
{\cal L}_0(|\bmu|),
\eqno{(4.3)}
$$
where
$
{\cal L}_0(|\bmu|)=(2\pi^3)^{-1/2}\delta^{-5}\,|\bmu|^2
$
is the total enstrophy when $\tau=0$.
In the linear problem
the normalized enstrophy ${\hat{\cal L}}$
depends only on the direction of $\bmu$.
Consequently, it can be put without loss of generality
that $|\bmu|=1$.

     Finally, for the normalized total enstrophy we obtain
$
{\hat{\cal L}}(\tau,\bmu)
=\ell_{ij}(\tau)\,\mu_i\mu_j.
$
\subsection{Enhancement (attenuation) of the vortex and
the relation to the hydrodynamic stability problem}
When $\tau\gg 1$ we obtain
$$
\ell_{ij}\approx l_{ij}\,\tau^2+O(\tau\ln\tau),
\eqno{(4.4)}
$$
$$
\left.
\ba{c}
l_{11}=\dfrac{9\upi^2+32}{288}\approx 0.41954,
\ \ l_{12}=\dfrac{5}{9}\approx 0.55556,\\
\\
l_{22}=\dfrac{9\upi^2-40}{36}\approx 1.35629,\ \
l_{33}=\dfrac{3\upi^2}{32}\approx 0.92529,
\ea
\right\}
$$
For the normalized enstrophy, when $\tau\gg 1$ we have
$$
{\hat{\cal L}}(\tau)\approx a({\bmu})\, \tau^2,\ \
a(\bmu)=l_{11}\,\mu_1^2+2\,l_{12}\,\mu_1\mu_2+
l_{22}\,\mu_2^2+l_{33}\,\mu_3^2.
\eqno{(4.5)}
$$
     By analyzing the coefficient $a(\bmu)$, we can readily
find the orientation of the vector $\bmu$ corresponding to those
initial vortices which will become {\it the most enhanced} at
large times. By fixing $|\bmu|=1$, we find that $a$ is maximum
when
$
\bmu=(\cos\alpha_{\infty},\sin\alpha_{\infty},0),
$
where
$$
\alpha_{\infty}={\frac{\upi}{2}}+{\frac{1}{2}}
\arctan\Bigl(\frac{2\,l_{12}}{l_{11}-l_{22}}\Bigr)\approx 65.07^{\circ}
$$
and is
$
a=a_{\rm \max}={\cas{1}{2}}\Bigl[(l_{11}+l_{22})+
\sqrt{(l_{11}-l_{22})^2+4\,l_{12}^2}\,\Bigr]\approx 1.6146.
$

     Note also that the angle $\sim 155.07^{\circ}$
with $a_{\rm min}\approx 0.1613$.
corresponds to the orientation of {\it the least
enhanced} (at large times) vortices, i.e. the normalized
enstrophy
of the vortices, the initial orientation angle of
which is directed along this direction, will be
an order of magnitude smaller than the maximum one.

     Further, it is assumed again that $\mu_3=0$.
In this case it will suffice to describe the orientation
of the initial vortex by only one inclination angle
of its fluid impulse
$\alpha$:
$
\mu_1=\cos\alpha,\ \ \mu_2=\sin\alpha,\ \ \mu_3=0.
$
In other words,
the angle $\alpha$ is the angle between the positive direction
of the $x$-axis and the direction of the initial fluid impulse $\bmu$.

     For an arbitrary time $\tau$ the maximum (over the
whole range of initial orientation angles $\alpha$)
enhancement of the enstrophy corresponds to the
initial inclination angle
$\alpha(\tau)$
%
$$
\alpha_{\rm opt}(\tau)={\frac{\upi}{2}}+{\frac{1}{2}}
\arctan\biggl[\frac{2\,\ell_{12}(\tau)}
{\ell_{11}(\tau)-\ell_{22}(\tau)}\biggr].
\eqno{(4.6)}
$$
A numerical calculation shows that $\alpha_{\rm opt}(\tau)$ increases
from
$\alpha_{\rm opt}(0)$ $=45^{\circ}$ to
$\alpha_{\rm opt}(\infty)=65.07^{\circ}$
\begin{figure}[t]
\epsfysize=40mm
\centerline{\epsfbox{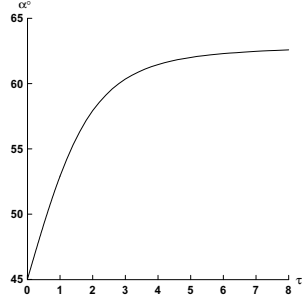}}
\caption{
The inclination angle $\alpha_{\rm opt}(\tau)$ of the
fluid impulse vector for which a maximum enhancement
of the enstrophy is reached by a given time $\tau$.}
\label{Fig. 5}
\end{figure}

     Figure 6 shows the normalized enstrophy as a function of
inclination angle $\alpha$ for
four values of $\tau$
$\tau=0$; 0.5; 1.0 and 5.0.
\begin{figure}[t]
\epsfysize=100mm
\centerline{\epsfbox{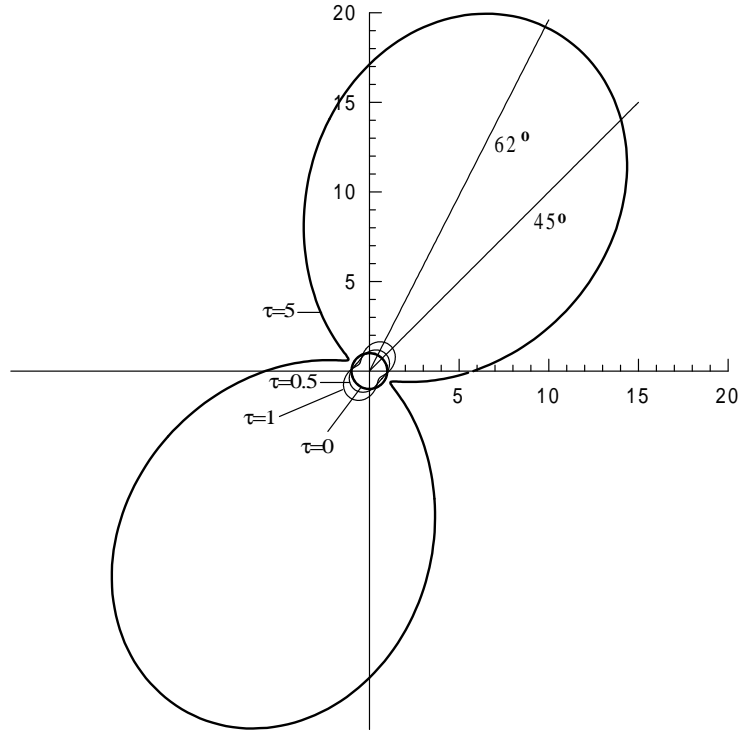}}
\caption{
The Normalized enstrophy (radial coordinate)
as a function of inclination angle of the
initial fluid impulse $\alpha$ (angular coordinate)
for four values of $\tau$: $\tau=0$; 0.5; 1.0 and 5.0.}
\label{Fig. 6}
\end{figure}

     The calculations show also
that at large $\tau$ almost all enstrophy is
concentrated  in horizontal directions of
vortex motion
${\cal L}_1\sim{\cal L}_3=O(\tau^2{\cal L}_2)$.
This fact is intuitively consistent
with results of numerical calculations of the isosurfaces
of enstrophy density (see \S\,3, as well as \S\,5) which
show that the localization plane of a {\it weak} vortex
becomes horizontal asymptotically.

As it follows from (4.5) for all possible orientation
angles of the initial vortex we obtain a power-law
(${\cal L}\sim \tau^2$)
growth of the total enstrophy (cf. with law of energy growth of the
optimal inviscid excitations in  F\&I).

     We may regard such an asymptotic increase in
enstrophy as the manifestation of an instability --
the initial
vortex starts to be enhanced infinitely (in linear theory at least).

     There is nothing surprising in this fact of a
power-law enhancement of enstrophy.
It is merely a reflection of another well-known fact,
namely, that Couette inviscid flow is unstable
with respect to 3-D
wave disturbances, and
this instability is just a power-law one, or,
as it is also called, an ``{\it algebraic}" instability
(Ellingsen \& Palm 1975).
Recall that Couette flow is stable with respect
to 2-D
wave disturbances ($k_3=0$) altogether (that is, there is
no not only the exponential instability but also
even the algebraic instability).

Note, that the last factor, together with the well-known
Squire (1933) theorem stating that if
the flow under consideration is stable with respect to 2-D
disturbances it is necessarily  stable with respect to
3-D disturbances), has long led to a paradox
laying  in the inconsistency between the experimental
fact of the existence of turbulence in Couette flow
and the absence of any instability in theory
(for more detail see an excellent review by Henningson, Gustavsson \&
Breuer 1994).
This paradox was just resolved by the discovery of the power-law
instability
($\omega_1(\tau,{\bf k})
\sim\omega_3(\tau,{\bf k})\sim \tau,\ \ \omega_2(\tau,{\bf k})\sim 1$)
of 3-D disturbances
caused by a so-called {\it lift-up} effect
(Landahl 1975)
which wasn't taken into account
in the proof of Squire's theorem.

     A localized vortex can be presented as a wave
packet composed of 3-D wave disturbances.
It is based on this that we have obtained here
a power-law increase of the total enstrophy of the vortex.


Precisely the same statement applies
for the evolution of a localized vortex on the
background of Taylor-Couette (circular)
flow considered by Malkiel, Levinski \&
Cohen 1999.
Here also  can not be an exponential growth of the
localized vortex in the range of parameters
where the hydrodynamic stability theory predicts a stability.
Now we have obtained the exact solution which describes the development of
the weak vortex in the circular flow. It would be presented in
separate paper.

     It is this fact that reflects the main conflict
of the theory L\&LC  with well-known facts
of the classical theory of hydrodynamic stability
mentioned in \S\,1.

     Note also that including the viscosity
is all the more unable to lead to an exponential growth
of the vortex. On the contrary, the presence of viscosity leads
to the fact that the vortex at some stage of its evolution can
cease to increase and subsequently begin to be dissipated
(see also F\&I). Moreover, a
sufficiently large viscosity
unavoidably lead to the {\it finiteness} of the lifetime
of a weak vortex. If, however, the viscosity is not too large and
the initial amplitude of the vortex is not too small, the
{\it nonlinearity} can come into play still before the vortex begins to
be decay. This issue requires the further (numerical, of course)
investigation.


\subsection{Enhancement of the enstrophy and Theodorsen's idea of the
predominant formation of 45-degree vortices}

     As early as five decades ago Theodorsen (1952)
came up with the hypothesis explaining why
the 45-degree direction of orientation of horseshoe vortices,
that is, the vortices whose plane is inclined at
$45^{\circ}$ to the basic flow direction, dominates
in experiments.

     Since Theodorsen's (1952) idea is most clearly
presented not in his work itself but in a later
publication by Head \& Bandyopadhyay (1981),
we shall follow the presentation of this paper
by adapting it to the present case of a linearized
problem and an external flow with $dU/dy=-\Omega={\rm const}$.

     We avail ourselves of the basic
equation of the theory (1.3a) omitting the nonlinear (underlined)
terms. A scalar multiplication of this equation  by $\bom$ gives
the equation describing the enstrophy dynamics of a fluid
particle:
$$
\frac{d}{dt}\bigl({\cas{1}{2}}\,\bom^2\bigr)=
-\Omega\,\omega_1\omega_2+
\Omega\left(\bom\,\frac{\p {\bu}}{\p z}\right)
+\nu\,(\bom\,\Delta\bom),
\eqno{(4.7)}
$$
where
$\dd/\dd t=\p/\p t+({\bU}\bnabla)$ is the Lagrangian time
derivative. The right-hand side of (4.7), with the exception of
the viscous term, is the linearized ``stretching term"
$\omega_i^t\,\omega_j^t\,\p u_i^t/\p x_j$, where the superscript
``t" corresponds to the total (non-linearized) value of the
physical quantity,
$\bom^t\equiv {\bOm}+\bom$,
${\bu}^t\equiv{\bU}+{\bu}$.

     By analyzing the first term on the right-hand side,
it is easy to see that it is maximal when the two-dimensional
vector $(\omega_1,\omega_2)$ (at its fixed absolute value) is
directed at the angle of $135^{\circ}$
to the positive direction of the axis $x$,
which corresponds to the angle of $45^{\circ}$ measured from
the direction of the mean velocity ${\bU}$ in the upper
half-space $y>0$. (We are reminded that $dU/dy<0$ corresponds
to positive values of $\Omega$.

     On the basis of this undeniable fact, Theodorsen
arrived to the conclusion that the concentration
of the vortex (its enstrophy) would also be maximal
in the plane oriented at the angle of $45^{\circ}$ to the flow,
and this does correspond to experimental findings.

     Over the course of the past five decades
Theodorsen's hypothesis has been repeatedly subjected
to criticism from different standpoints.
Here we want to discuss only two aspects, based on
the just obtained (on the basis of the {\it exact} solution)
results on the evolution of total enstrophy,
results of calculations of the 3-D
vorticity field presented in \S\,3, as well as on
the results of calculations of the vortex localization
plane inclination which will be described later in  \S\,5.

     The first aspect implies that the
conclusion about the greatest buildup rate of 45-degree
vortices was drawn, strictly speaking, from
analyzing the structure of only one of the terms on the
right-hand side of (4.7). Our analysis shows, however,
that the second term,
$\Omega\,(\bom\,\p{\bu}/\p z)$,
that represents the
other part of the ``stretching term" responsible
for the distortion of the flow velocity field
caused by the vortex is also important.
Although the integral contribution of this term
to the total enstrophy is exactly zero
at the {\it initial} instant of time, it becomes
substantially larger with the time
and can compete with the integral contribution of the first term.

     Thus the obviously true statement about the
role of the stretching term made by Theodorsen can be applied,
strictly speaking, only to  the initial instants of time, $\tau\ll 1$,
when the distortion of the flow velocity field
still can be neglected.  Which one of
the initial vortices will turn out to be the
most enhanced for a sufficiently large time,
in the course of which the direction of velocity
is significantly changed, now becomes quite unclear
from the  reasoning presented.

     And the second aspect, which is of course
associated with the first one, implies that at
this point, taking into consideration the change of the
velocity field orientation (actually neglected
in Theodorsen's discussion), it becomes totally unobvious
that the initial vortex with the optimal $45^{\circ}$-orientation,
which {\it at the initial instant of time} was
enhanced faster than all the others, would not change
the orientation of its plane in the course of evolution.

     From the results presented in \S\,4.2
(remember that they refer to the inviscid case) that are most
instructively illustrated in figures 5 and 6, it follows that at
small $\tau$ the 45-degree vortices are indeed the strongest.
However, at larger times,
quite different vortices turn out to be most strongly enhanced,
i.e. those for which at the initial instant of time the localization
plane was more strongly pressed against the flow direction). In
the limit $\tau\gg 1$ the vortices, which initially were inclined
at $90^{\circ}-\alpha_{\rm opt}(\tau=\infty)=25^{\circ}$ rather than
$90^{\circ}-\alpha_{\rm opt}(\tau=0)=45^{\circ}$ turn out to be the
strongest.

     It should be noted at this point that,
if the relatively small
differences between  the predicted
angles of maximum enhancement are not taken
into consideration, Theodorsen's hypothesis is plausible enough.

As can be shown that including of
the viscosity is also in favor of this hypothesis.
At a finite Reynolds number, $Re=\Omega\,\delta^2/\nu$, the
difference between the angle of maximum enhancement and the angle
of $45^{\circ}$, as predicted by Theodorsen, becomes still
smaller. This can be most easily understood from the expression
for total enstrophy at small $\tau$  obtained with the including of
viscosity (the details of its derivation are
omitted):
$$
{\hat{\cal L}}(\tau,\alpha;Re)
=1+{\cas{1}{2}}\tau\Bigl[\,\sin(2\alpha)-
\frac{20}{Re}\,\Bigr]+O\left(\tau^2\right).
\eqno{(4.8)}
$$
It follows from (4.8) that  initially the
45-degree (or, what is the same, 225-degree) vortices are the
strongest. And the vortices whose plane is inclined at
$135^{\circ}$ are, on the contrary, the weakest (they are even
weaker than the initial vortex, i.e.
${\hat{\cal L}}<1$).
It turns out that in the case of a sufficiently large viscosity the
angles that are only very close to $45^{\circ}$  are enhanced, and
when
$Re=Re_{\rm cr}=20$
the only direction, $45^{\circ}$, is
enhanced altogether. Calculation of the total
enstrophy with the including of viscosity, shows
that vortices of other directions, close to $45^{\circ}$, will
start to be enhanced with the passage of time,
however, the angles of maximum enhancement
remain close around $45^{\circ}$ during all time up to
the beginning of the dissipation of the vortex.

     For illustration we presented in Fig.~7 the contours
of ${\hat{\cal L}}(\tau;\alpha)$ for viscous case with
$Re=20$ and $Re=40$.
The shaded regions of the plane $(\tau,\alpha)$ correspond to
${\hat{\cal L}}(\tau;\alpha)>1$, i.e. to the
enhancement of the vortex, and the unshaded areas
correspond to ${\hat{\cal L}}(\tau;\alpha)<1$,
that is, to the attenuation of the vortex.
We see that if $Re=20$ the vortex begin to dissipate
for all $\alpha$ (except $\alpha=45^{\circ}$ and $225^{\circ}$),
although with the growth of time
some orientations near $45^{\circ}$ also (very weakly) enhance.
\begin{figure}
\epsfxsize=100mm
\centerline{\epsfbox{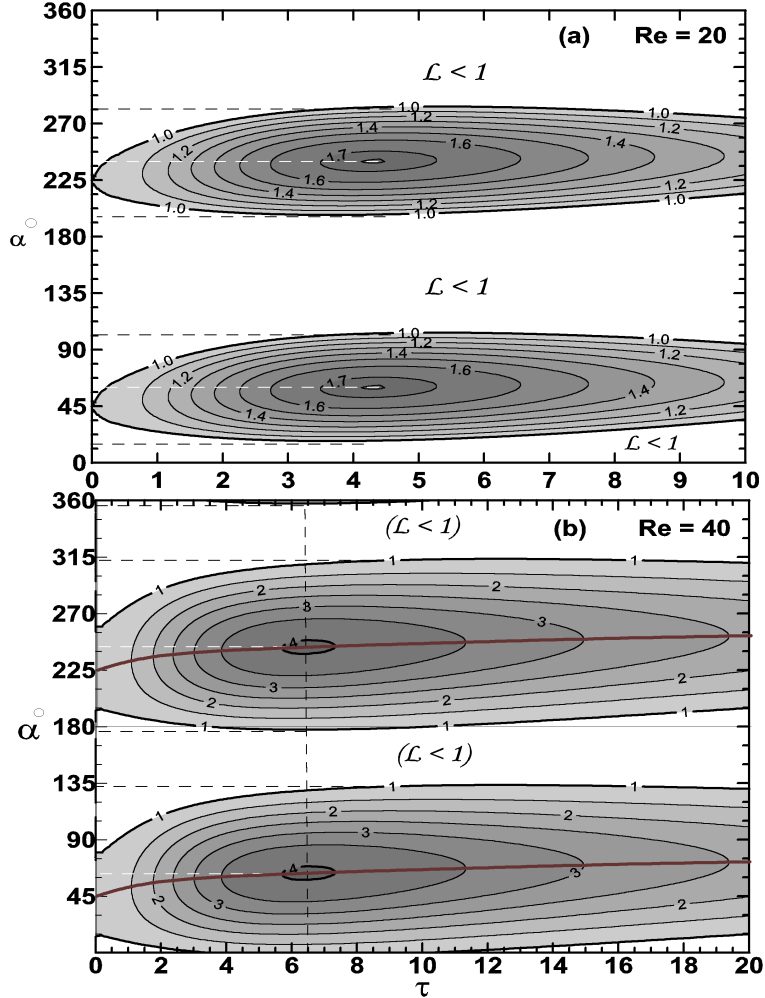}}
\caption{
Contours of the normalized
enstrophy ${\hat{\cal L}}(\tau;\alpha)={\rm const}$
on the plane $\tau-\alpha$  for $Re=20$ and $Re=40$.
 Unshaded areas
correspond to the attenuation of the vortex, ${\hat{\cal L}}<1$.
}
\label{Fig. 7}
\end{figure}

It is interesting to note
that this critical Reynolds number is very close to
the value $Re_{\rm min}=2\pi^2\approx 19.7$ obtained in F\&I
for growth of energy of optimal checkerboard excitations
with $k_1=0$.

     The situation with the second point concerning
the inclination angle of the localization plane
of the vortex is much less favorable.
If we keep track on the evolution of the inclination
angle of the enstrophy localization plane (vortex plane),
we find that at large times this angle {\it tends
to zero} (i.e. the plane of the vortex tends to become horizontal),
rather than to $45^{\circ}$.
This statement is illustrated in figure 4,
as well as by calculations of this angle
performed on the basis of the tensor
of enstrophy distribution (TED) which are presented in \S\,5.

\section{The tensor of enstrophy distribution and vortex geometry}
     We can often avoid an unwieldy description
of the vortex by specifying its total vector
field if we are able to introduce some integral
characteristic of the vortex which (although not reflecting,
of course, in full measure the entire vector
structure of the vortex) will permit its main geometrical
parameters to be described at least roughly.

     For this purpose we avail ourselves
of the analogy with those approaches which are
used in electrostatics in describing the
distribution of electric charge. This is customarily
done by using so-called multipole moments.

     We now introduce the notion of the
Tensor of Enstrophy Distribution, TED, which
is essentially a usual {\it quadrupole moment}
of the enstrophy distribution.

     We assume to use the following definition of the tensor
(see, for example, the book by Levich 1969):
$$
T_{ij}=\int \dd V \bom^2({\br})\,x_i x_j.
\eqno{(5.1)}
$$
Note, that if the displacement
of center of vorticity distribution takes place
(i.e. if
$
X_j\equiv \left(\int dV \omega^2 x_j\right)/
\left(\int dV \omega^2\right)\ne 0$,
as it may be in the case of strong vortices) the definition of TED
must be generalized:
$T_{ij}=\int \dd V \bom^2({\br})\,(x_i-X_i)(x_j-X_j)$.

As any symmetric tensor, it can be transformed to the
principal axes, $x_i'$, where it has a diagonal form:
$$
{\hat T}'=\left|\,\,
\ba{rcl}
\lambda_1&0&0\\
0&\lambda_2&0\\
0&0&\lambda_3
\ea
\,\,\right|,
\eqno{(5.2)}
$$
Here $\lambda_i$ stands for the eigenvalues
of the matrix $T_{ij}$, i.e. the solution of a characteristic equation
$
{\rm Det}\,||T_{ij}-\lambda\delta_{ij}||=0.
$
Then the direction of one of the principal
axes that corresponds to the {\it smallest}
of these three values of $\lambda$, is the direction
which should be identified with a normal to the
vortex plane. And the vortex itself is
extended along the direction which
corresponds to the {\it largest} value of $\lambda$
(see Fig.~8).
\begin{figure}
\epsfysize=55mm
\centerline{\epsfbox{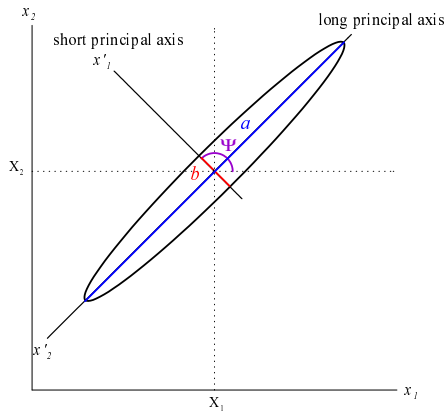}}
\caption{
Illustration to the explanation of TED.}
\label{Fig. 8}
\end{figure}

     It is also possible to introduce
the notion of the size of the vortex $a_i$ along
the corresponding principal axes $x_i'$:
$$
a_i=\sqrt{\frac {T_{ii}'}{\cal L}}=\sqrt{\frac{\lambda_i}
{\int \dd V \bom^2({\br})}.}
\eqno{(5.3)}
$$

     Let the largest axes be referred to as $a$,
the smallest as $b$, and let the third axis be $c$.

     For illustration of the meaning of TED
let us consider the initial Gaussian vortex. We have in this case:
$
T_{ij}=K\,(2\mu^2\delta_{ij}-\mu_i\mu_j)\equiv
 K\,t_{ij},
$
where
$
K=1/(4\sqrt{2}\upi^{3/2}\delta^3)
$.
It is easy to see (transforming TED to the principal axes)
that the direction along $\bmu$ does correspond
to the smallest of the three values of $\lambda$, i.e. $\mu^2$,
and in the plane which is normal to this direction,
the eigenvalues are identical and are $2\mu^2$. The size
ratio along $\bmu$ and in the plane
perpendicular to $\bmu$ is $1:\sqrt{2}$.

     Thus the TED characterizes rather
substantively the distribution of enstrophy
for which we have an instructive visual idea from figure~1.

     Further, we again restrict our discussion to
the case of a symmetric (about the plane $z=0$) vortex, $\mu_3=0$.
In this case it is easy to see that the axis $x_3$ (axis $z$)
remains one of the principal axes over the course
of the entire evolution, and the TED
has a more straightforward form
$$
{\hat T}=
\left|\,\,\,
\ba{ccc}
A&C&0\\
C&B&0\\
0&0&D
\ea
\,\,\right|.
\eqno{(5.4)}
$$
In this case the tensor is transformed to the principal
axes by a simple rotation of the plane $(x_1,x_2)$ around
the axis $x_3(=x_3')$ by an angle $\bPsi$
$$
\tan 2\bPsi=\frac{2\,C}{A-B},
\eqno{(5.5)}
$$
and has in these axes a diagonal form.
Let the axis $x_1'$ coincide with the shortest
principal axis. Then in the new axes we obtain
$$
{\hat T}'=\left|\,\,\,
\ba{ccc}
{\cas{1}{2}}(A+B-S)&0&0\\
0&{\cas{1}{2}}(A\!+\!B\!+S)&0\\
0&0&D
\ea
\,\,\right|,
\eqno{(5.6)}
$$
where $S=\sqrt{(A-B)^2+4C^2}$
and the angle between the positive direction
of the axis $x$ and the direction of a normal to the
plane of the vortex (i.e. the axis $x'(\equiv x_1')$)
$$
\bPsi={\cas{1}{2}}
\arctan\Bigl(\frac{2C}{A-B}\Bigr)+\cas{1}{4}\upi\,(1+s),\ \
s={\rm sign}\,(A-B).
\eqno{(5.7)}
$$

     Next, using the notion of the TED introduced above,
we can employ it to calculate the geometrical characteristics of
the vortex and compare them with results that follow from
calculations of the 3-D vorticity field by exact
formula (3.5).

     Results of calculations of the vortex parameters,
obtained on the basis of the TED, for four
initial directions of the vector
$\bmu$, $\alpha=0^{\circ},\ 45^{\circ},\ 90^{\circ},\ 135^{\circ}$
(where $\bmu = (\cos\alpha,\,\sin\alpha,0)$)
are shown in Fig.~9.
\begin{figure}
\epsfxsize=110mm
\centerline{\epsfbox{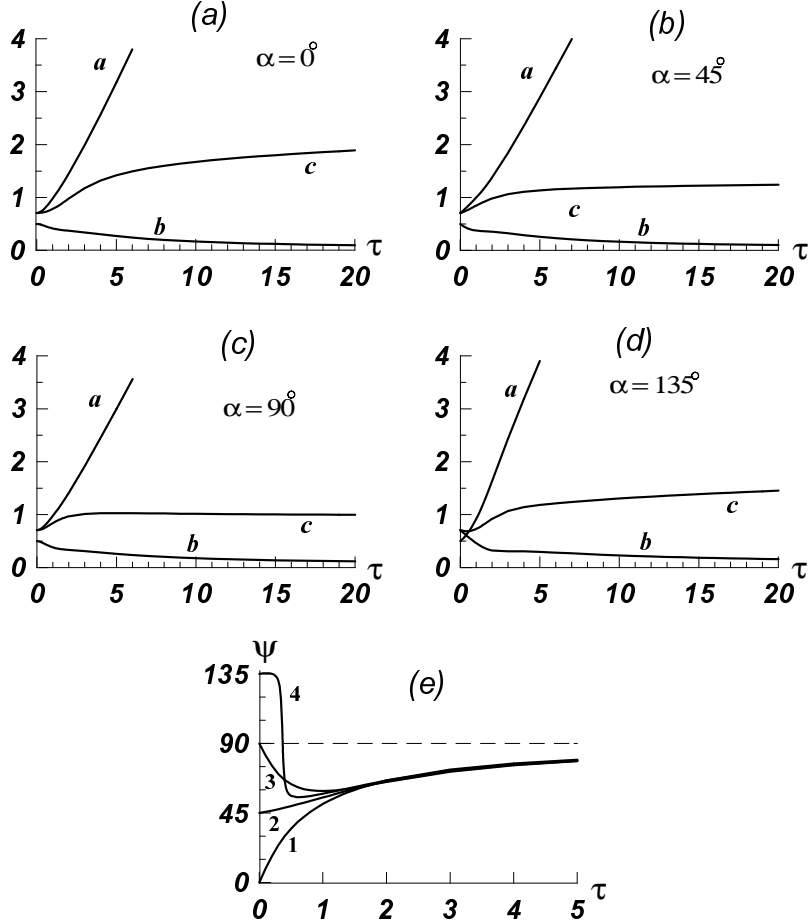}}
\caption{
The temporal dependence of the vortex parameters.
The ``axes" $a$, $b$ and $c$
for the initial values of the inclination angle $\alpha$:
$(a)$ -- $\alpha=0^{\circ}$, $(b)$ -- $45^{\circ}$,
$(c)$ -- $90^{\circ}$, $(d)$ -- $135^{\circ}$ and
$(e)$ --  the inclination angle $\Psi$
as a function of $\tau$ for these four initial values.}
\label{Fig. 9}
\end{figure}

     It is evident that at sufficiently large $\tau$ the
normal to the plane of the vortex is nearly vertical,
$
\bPsi\approx \upi/2-1/\tau.
$

     In order to estimate the effectiveness of TED
with respect to description of the vortex geometry,
the parameters of the vortex geometry obtained
from TED calculations were compared with results following
from calculations of the complete vorticity field
by exact formula (3.5).

     An exellent agreement of the inclination anlges $\bPsi$
following from the TED  with the actual inclinaton angles
of the  planes of enstrophy localisation
was shown by Suponitsky, Cohen \& Bar-Yoseph (2003, 2004).
%

     Thus it can be stated that the TED is
a rather convenient and reliable integral characteristic for the
description of the vortex dynamics. Of course, it is unable to
describe the vector structure of the vortex, yet it can be used
to obtain a sufficiently great deal of information about the
vortex.

%

     The results on the orientation
of vortex plane, obtained in this paragraph, unlike the findings concerning
the evolution of total enstrophy, described in \S\,4
are virtually insensitive to the presence of viscosity.
(It follows from our calculations of TED in viscous case,
not presented here.)

%
\section{Discussion}
     As has been pointed out in the \S\,1,
the motivation for this investigation was the analysis
of a theoretical model suggested by Levinski (1991)
for explaining the evolution mechanism of
localized vortices observed in turbulent boundary layers. A
key point in this model implies separating
the complete vorticity field into the concentric
vorticity with vortex lines enclosed within the region
immediately surrounding the initial vortex disturbance,
and the vorticity field associated with vortex
``tails" which are produced in the process
of evolution of the initial vortex disturbance.
It should be noted that the possibility of such
a separation is not strictly substantiated
mathematically but is accepted in Levinski (1991)
and in subsequent publications
(Levinski \& Cohen 1995; Levinski, Rapoport \& Cohen 1997;
Malkiel, Levinski \& Cohen 1999 and Levinski 2000)
as a physically justified hypothesis.
The criterion of correctness of this approach
comes from the agreement between results of
theoretical analysis and experimental results.
In particular, the predictions obtained on the basis of the model
suggested in L\&LC for rotating Couette
flow were confirmed experimentally by Malkiel, Levinski \& Cohen (1999).

     In this paper the hypothesis about the
possibility of separating the vorticity is verified by
constructing the complete vorticity field at an arbitrary instant
of time for a small  amplitude localized disturbance.
The problem of the evolution of a weak localized disturbance is
analyzed on the basis of {\it exact} solution
for the external constant shear flow
To ease the subsequent analysis of
the vorticity field, the initial vortex disturbance was
represented by ``Gaussian vortex" (2.15) that specifies a very
simple localized vortex, having the structure of a vortex dipole.

     It has been shown that in accordance with
the classical stability theory results
(see, for example, Dikii 1976), the vorticity
amplitude increases not faster than it does as a power-law.
This result {\it contradicts} the exponential growth of the fluid
impulse obtained in L\&LC
for the ``core" of the vortex disturbance identified in
a special way. This could be accounted for by the fact
that the generation of a new vorticity in the process
of evolution of the vortex disturbance can lead to a fast
increase of the ``mass" of the vortex ``core".
It is this phenomenon that is observed in visualizing
vortex structures in turbulent boundary layers.
Specifically there is a rapid growth
(in the sense of the {\it geometrical} growth) of hairpin
vortices which represent localized vortex dipoles.

     In order to analyze this possibility,
we introduce the notion of the {\it modified fluid impulse} (MFI)
defined as an integral of the dipole moment of vorticity
over the infinite spherical volume.
This definition coincides formally with the
definition of the fluid impulse used in L\&LC,
but, unlike the latter, it is defined
for the {\it complete} vorticity field. In doing this, we,
using only the property of the localized character
of the disturbance and without imposing constraints on its
amplitude, show that if the MFI exists at the initial
instant of time, then it exists also at any subsequent
instant of time and does not depend on the particular
coordinate system chosen.

     An analysis of the MFI behavior over large times
$t \gg 1/|\Omega|$ shows that for any initial vortex
orientation
the MFI increases {\it not faster than linearly} with the time.

     This result enables us to verify in a direct
manner the hypothesis
proposed in L\&LC concerning the possibility of
separation of the enclosed concentrated vorticity
(localized vortex ``core")
from the complete disturbed  vorticity field. Indeed, let
us assume that such a separation is possible, that is,
$\bom=\bom^I+\bom^{II}$, where $\bom^I$ describes the vortex
``core", and $\bom^{II}$ describes the vortex ``cloud" that
includes all ``tails" of the complete vorticity field. For each
vorticity field, one can determine, in accordance with the
expression (2.4), its modified fluid impulse, so that
${\tilde{\bp}}={\tilde{\bp}}^I+{\tilde{\bp}}^{II}$
Furthermore, as in L\&LC, ${\tilde{\bp}}^I$ is a {\it
true} fluid impulse. Accordingly, equation (2.10) that describes
the dynamics of MFI defined for the complete vorticity field, breaks down
into two equations
$$
\frac{d {\tilde p}^I_i}{dt}=
-{\cas{1}{2}}\,{\tilde p}^I_j\,\,\frac{d U_i}{dx_j}-
{\cas{1}{2}}\,{\tilde p}^I_j\,\,\frac {dU_j}{dx_i},
\eqno{(6.1)}
$$
$$
\frac{d {\tilde p}^{II}_i}{dt}=
-{\cas{1}{2}}\, {\tilde p}^{II}_j\,\,\frac{dU_i}{dx_j}
-{\cas{1}{2}}\,{\tilde p}^{II}_j\,\, \frac{dU_j}{dx_i}
+\lim\limits_{R\to\infty} J_i(R).
\eqno{(6.2)}
$$
Note that, according to the separation condition, the
``tails" of the complete vorticity field make a contribution to
the dynamics of ${\bp}^{II}$ only.

     An exponential growth of ${\tilde{\bp}}^I$
follows from equation (6.1).
The fact that the sum
${\tilde{\bp}}^I+{\tilde{\bp}}^{II}$
increases  not faster than as a
power-law, implies that ${\tilde{\bp}}^{II}$ also increases
exponentially fast. It should be noted here that the term $\bJ$  in
equation (6.2), describing the contribution from
the vorticity ``tails"
to the dynamics of ${\tilde{\bp}}^{II}$ grow
also {\it not faster} than as a power-law. Thus the main
contribution to the MFI dynamics for the field $\bom^{II}$ is
made by the region that immediately surrounds the vortex ``core"
and, hence, the {\it assumption about the possibility of
separating the vortex core is invalid}.

     On the other hand, the fluid impulse, defined for
the complete vorticity field, cannot be an adequate
characteristic of the evolution of a localized vortex. The formal
reason is the fact that the volumetric integral involved in the
definition of the fluid impulse is not absolutely convergent, and
its value depends on the form of the integration domain when its
size is made tend to infinity. In this paper this is illustrated
by a comparison of the asymptotic values of the fluid impulse for
two cases where the region of integration represents a spherical volume,
first in Euler coordinates, and then in Lagrangian coordinates.

     In summarizing all attempts to describe the
evolution of a localized vortex in the external shear flow,
it can be stated that using the {\it moments of the vorticity
field} in this problem is unjustified.

     In order to be able to describe the enhancement
or attenuation of the vortex and the variation of
its orientation the course of  the evolution,
we have analyzed the evolution  of the total enstrophy
of the vortex (4.1) and of the tensor of enstrophy distribution (5.1).
They permit the evolution of the vorticity
amplitude and the main geometrical characteristic
of the vortex to be described by means of only
a few independent parameters.

     In particular, the effectiveness of the
description of the vortex on the basis of the tensor  of enstrophy
distribution (TED) can be demonstrated by comparing
visual pictures of enstrophy density isosurfaces
$|\bom({\br})|^2={\rm const}$,
constructed on the basis of
the exact solution for the complete vorticity field,
with what follows from the description based on TED
for the inclination angles of the plane of the vortex.
Thus the TED is a reliable alternative
(to the fluid impulse) integral characteristic
that enables an instructive
representation of its evolution, instead of an unwieldy
description using the complete vorticity field.

Thus, the calculations done in this paper show
that the linear evolution results in a pair
of rollers lying in the horizontal plane
and aligned along the flow. It is interesting
to note that this result is in excellent
agreement with the finding reported
in F\&I where the evolution of the form of
energetic isosurfaces was calculated
for optimal checkerboard perturbation.
In spite of the large difference of
the initial perturbations considered
in our paper and in F\&I, the outcome of their
evolution turned out to be strikingly
alike (see figures 9 and 10 in F\&I).
This indicates the universal
character of the mechanisms giving rise
to ordered structures in the course
of the evolution of 3-D perturbations in shear flows,
as declared in F\&I, and also lends support
to the idea that  the particular form of the initial
vortex selected in our paper is not very important.

However, these results that follow from
the exact solution of the evolution problem for
the small amplitude localized vortex, {\it contradict} the known
experimental facts obtained by visualizing hairpin vortices
developing in turbulent boundary layers (Head \& Bandyopadhyay 1981)
or artificially synthesized in laminar boundary layers
(Acalar \& Smith 1987\,{\it a,b}).

     Moreover, as shown in \S\,3,
the symmetry properties of the basic
equations in the linear case, in principle,
do not allow the formation of hairpin vortices.
For that reason, (numerical) investigation of
the nonlinear stage of evolution of a
localized vortex is of utmost current importance.
Preliminary results of numerical simulations
with strong vortices confirm the occurrence of hairpins
at a certain stage of vortex evolution
(Suponitsky {\it et al}, 2003, 2004).

\medskip
\acknowledgments

We express our profound appreciation
to V. Suponitsky, Y. Cohen and P. Bar-Yoseph
for useful collaboration and to S. M. Churilov and A. M. Fridman
for encouragement and helpful discussions.
Thanks are also due to Mr V.~G.~Mikhalkovsky for
his assistance in preparing the English version of the manuscript.

\end{document}